\documentclass{aa}
\usepackage{epsfig,amssymb}
\def \gg{$\gamma$-ray}
\def\scri#1{\mbox{\scriptsize #1}}
\def\sc#1{_{\mbox{\scriptsize #1}}}
\def\mc{M$^{\scri{te}}$\,Carlo}
\def\be{\begin{equation}}
\def\ee{\end{equation}}
\def\dh{\mathcal{H}}
\def\ch{\mathrm{H}}
\def\dm{\mathcal{M}}
\def\cm{\mathrm{M}}
\def\dw{\mathcal{W}}
\def\cf{\mathrm{F}}
\def\gl#1{Eq.\,\ref{#1}}
\def\ab#1{Fig.\,\ref{#1}}
\def\3{ss}

\renewcommand{\today}{\,}
\begin{document}

\thesaurus{03 
              (03.13.2;  
               03.13.6)} 
   \title{A fast unbinned test on event clustering in Poisson processes}

   \subtitle{}

   \author{J. Prahl}

   \offprints{J. Prahl, prahl@mail.desy.de}

   \institute{II. Institut f\"ur Experimentalphysik, Universit\"at Hamburg,
              Luruper Chaussee 149, D-22761 Hamburg\\
              email: prahl@mail.desy.de}

   \date{Received ???; accepted ???}

   \maketitle

   \begin{abstract}

   An unbinned statistical test on cluster-like deviations from
   Poisson processes for point process data is introduced, presented 
   in the context of time variability analysis of astrophysical
   sources in count rate experiments. The measure of deviation
   of the actually obtained temporal event distribution from that of a Poisson
   process is derived from the distribution of time differences between two
   consecutive events in a natural way. The differential character of the
   measure suggests this test in particular for the search of irregular
   burst-like structures in experimental data. The construction allows the
   application of the test even for very low event numbers. Furthermore, the
   test can easily be applied in the case of varying acceptance of the
   detector as well. The simple and direct use of background events
   simultaneously acquired under the same conditions to account for
   acceptance variations is possible, allowing for easy application
   especially in earth-bound \gg\ experiments. Central features 
   are the fast and simple calculation of its measure, and the existence 
   of an analytical approximation that describes the general test 
   statistics to a high degree of precision.
      \keywords{methods: data analysis -- methods: statistical}
   \end{abstract}

%

\section{Introduction\label{phinf}}
   In the search for high energy astrophysical \gg\ sources, 
   be it with earth-bound experiments or with satellite borne 
   detectors, the test on an integral (DC) excess of event numbers
   from the direction of a source, compared to an adequately derived 
   background expectation, is certainly the major statistical tool.
   (For a quite general consideration of this topic,
   see Li\,\&\,Ma \cite{lima}).
   However, an integral excess of events is not the only
   characteristic one might be interested in. Many \gg\ sources are known
   to exhibit highly variable fluxes. 

   In this paper a newly developed test on variability for photon counting
   experiments is presented. 
   In contrast to time series analyses, in which sampled continuous 
   functions are examined and which are applied e.\,g.\ in optical astronomy,
   this test deals with discrete registration times of single
   events, i.\,e.\ {\em point process data}. This situation is typically
   encountered in \gg\ astronomy.

   It is dedicated to two basic, different, but closely related
   questions that require statistical methods:
   \begin{itemize}
   \item[1.] Are there significant signals of variability from a certain 
   celestial position even if there is only a moderate or hardly significant 
   DC excess from that direction, if the observation is limited by
   background fluctuations?
   \item[2.] How well is the sequence of arrival times of pure (or almost pure) 
   events from a well-known source described by a constant flux?
   \end{itemize}
   Considering the case of known or assumed periodicities of activity of 
   the objects in question, several statistical tests are established to
   search for fixed frequencies in point process data 
   (see e.\,g.\ Mardia \cite{mard}, Lewis \cite{lewis}). 
   
   For the remaining case of irregular temporal activity schemes, the
   situation concerning established tests is less
   satisfying. In many publications, more or less spontaneously invented
   measures of variability are used, based mostly on a procedure of
   binning the event data in equally sized time slices and subsequent
   application of a $\chi^2$ test.

   Based on this idea, some rather elaborated tests have been worked
   out, binning the data into a lot of different ways under variation of
   bin phases and widths (e.\,g.\ Collura et\,al.\ \cite{coll},
   Biller et\,al.\ \cite{bill}). The measure of variability is then
   obtained by a certain combination of all single $\chi^2$ test results.
   Such tests typically suffer from two deficiencies:
   \begin{itemize}
   \item[1.] The chance probability for a given probe can in general only
   be assessed  by \mc\ means, since the single $\chi^2$ test results are
   not independent
   \item[2.] The computation of the measure itself is quite 
   expensive expressed in terms of computing time and storage needs
   since the data has to be divided into a lot of different binning schemes
   \end{itemize}
   Other drawbacks in certain situations are the restriction of those tests
   to sufficiently high event statistics, and the fact that the rigorous
   employment of simultaneously acquired background events to account for
   a non-uniform temporal acceptance of a detector 
   is not obvious. In addition, the procedures are by far too complicated
   to apply them to a huge number of different potential source directions,
   e.\,g.\ using them for a kind of all-sky search for variable sources.

   It should not be concealed that there exist alternative
   variability analysis procedures for point process data (see e.\,g.\
   Scargle \cite{scarg} or Gregory\,\&\,Loredo \cite{greg} for an
   application of Bayesian methods). These are surely appropriate to
   attack ``higher order questions" such as typical time scales of 
   variability or change points of fluxes, but the application is at least as
   complicated as the use of binned tests.

   These difficulties may well bring down the willingness of 
   experimenters to apply those tests.   
   The unbinned {\em exp-test} that will be introduced in the following is,
   in contrast to the tests mentioned above, easy to apply,
   using a straightforward, natural measure for variability.
   The result of it will therefore not be as specific as for most of
   the above tests, but will supply a measure for variability in general.

   It should be emphasized that one established test of this kind is already 
   existing: the nonparametric Kolmogorov test, applied to the cumulated 
   distribution function (cdf) of the registration times of the
   events in question, 
   comparing them to the expected equal distribution.
   In opposition to it, the exp-test will have a more differential
   character, thus complementing the Kolmogorov test in a
   sense that will be specified 
   later. 

   Although described here in the context of time variability analysis,
   the application of the exp-test is not restricted to the time domain,
   but may be used in the a\-na\-ly\-sis of e.\,g.\ spatial or frequency data 
   as well. It might also be found useful in other disciplines
   of science where the analysis of empirical data plays
   a role, like biology, economics and engineering.
   
   In the following section, the exp-test is developed for the ideal
   situation of a uniform or exactly known temporal acceptance function
   of the experiment with respect to the events from a potential source.
   Section 3 deals with the generalization needed to apply the exp-test
   with simultaneously registered background events, while in section
   4 the sensitivity of the test is characterized. In section 5, a brief
   comparison with the Kolmogorov test is carried out, and in the last
   section a summary is given.

\section{Developing the exp-test}
Supposed a uniform temporal acceptance of the used detector is given,
the expected temporal sequence of registered events in absence of a variable 
source is governed by Poisson statistics. This applies to both background
events and events from a steady source, thereby defining the zero 
hypothesis for {\em any} test on variability in this context.

One starts with the Poisson distribution
\be
P_\lambda(n) = e^{-\lambda}\cdot\frac{\lambda^n}{n!}
\label{poisson}
\ee
where $\lambda$ denotes the expectation value.\\
A monotone sequence $(t_i)$ of events in time $t$ is called a
{\em Poisson process}, if there exists a constant $C$, such that for all
$\Delta t>0$, dividing the time in equally sized intervals of the length
$\Delta t$, the numbers of events per time interval are 
Poisson distributed with $\lambda = \Delta t/C$ and mutually independent
(see e.\,g.\ Chatfield \cite{chatf}). 
From this immediately follows the probability density function
(pdf) of the time intervals $\Delta t$ between two consecutive events
as the derivative of $P_\lambda(0)$ with respect to $\Delta t$
\begin{eqnarray}
\nonumber 
f_C(\Delta t) & = &
-\frac{P_{(\Delta t + d\Delta t)/C}(0) - P_{(\Delta t)/C}(0)}{d \Delta t} \\ 
  & = &-\frac{d\exp(-\Delta t/C)}{d \Delta t}
   = \frac{1}{C}\cdot \exp\left( -\frac{\Delta t}{C}\right)
\label{dpoisson}
\end{eqnarray}
i.\,e. an exponential distribution in $\Delta t$ with the mean value $C$.\\ 

Let there now be a randomly chosen time interval, containing the 
monotonically increasing sequence of event times
$(T_i)_{i=1\ldots N+1}$. Let the resulting distribution 
\be
\lbrace\Delta T_i\rbrace _{i=1\ldots N} :=
  \lbrace (T_{i+1}-T_i)\rbrace_{i=1\ldots N}
\ee
have a mean value  $\overline{\Delta T} =: C^\ast$. 
If one sets $C=C^\ast$ in \gl{dpoisson}\ (i.\,e.\ setting $C$ to the 
actually obtained mean value), it follows for the $\Delta T_i$ to stem
from a distribution $f_{C^\ast}(\Delta t)$, under the constraint of the
conserved mean value. This is generally valid, even for {\em a priori} 
unlikely values of $C^\ast$, because the $T_i$ are completely uncorrelated.
This ensures that any analysis based only on $C^\ast$ is
independent of a possibly present DC excess or deficit.

Now the distribution of the $\Delta T_i$ can be compared with
$f_{C^\ast}(\Delta t)$ using a distribution test. The classical tests 
(e.\,g.\ Kolmogorov test or Smirnov--Cram\'er--von-Mises test) are neither
well suited to measure exactly the effect that is searched for, nor is it
possible to take the constraint of the conserved mean value into account.
For this reason a natural measure will be introduced here that is 
especially sensitive to excesses of the $\Delta T_i$ far from the mean value.

Defining the pdf $\cf(\Delta t)$ of the random probe 
\be
  \cf(\Delta t):= \frac{1}{N}\sum_{i=1}^N \delta(\Delta t - \Delta T_i)
\ee
the equalities of the 0.\ and 1.\ momenta of $f_{C^\ast}$ and $\cf$ serve as a
starting point:
\begin{eqnarray}
 \int\limits_0^\infty f_{C^\ast}(\Delta t)\, d\Delta t 
           \enspace =& {\displaystyle\int\limits_0^\infty} \cf(\Delta t)\,
               d\Delta t & = \enspace 1 \label{poismom1}\label{zerothmom}\\
 \int\limits_0^\infty \Delta t\cdot f_{C^\ast}(\Delta t)\, d\Delta t
 \enspace =& {\displaystyle\int\limits_0^\infty} 
  \Delta t\cdot \cf(\Delta t)\, d\Delta t & = \enspace C^\ast
\label{poismom2}
\end{eqnarray}
The variance of the distribution $\cf(\Delta t)$ could be used as a measure
for event clusters (and corresponding dilutions), but this quantity is
too sensitive on very large $\Delta T_i$ due to the extreme assymmetry of
the exponential distribution. (Actually the value of the variance 
is dominated by the pdf between $C^\ast$ and $\infty$, although only
37\,\% of the $\Delta T_i$ are contained in that interval.)
In order to be not too sensitive to outliers, another way will be followed.
From Eqs.\,\ref{poismom1},\,\ref{poismom2} follows
\begin{eqnarray}
\nonumber
  \lefteqn{
  \int\limits_0^\infty \underbrace{\left(1-\frac{\Delta t}{C^\ast}\right)}_{
  \displaystyle := h(\Delta t)} 
    \cdot f_{C^\ast}(\Delta t) \,d\Delta t}\\
  & & \displaystyle = \int\limits_0^\infty 
\left(1-\frac{\Delta t}{C^\ast}\right)\cdot \cf(\Delta t) \,d\Delta t =0
\label{t01}
\end{eqnarray}
Defining now 
\be
 \ch (x) := \int\limits_0^x h(\Delta t)\cdot f_{C^\ast}(\Delta t)\, d\Delta t 
\label{hvonx}
\ee
it can be concluded
\begin{itemize}
\item $\ch (0) = \ch (\infty) = 0$
\item $\ch (C^\ast) = 1/e$ ist the global maximum of $\ch$, as $f_{C^\ast}$
      is positive everywhere, and $h(\Delta t)$ is
      monotonically falling, being zero at $C^\ast$
\end{itemize}
Replacing in \gl{hvonx}\ $f_{C^\ast}$ by $\cf$, the first property
holds, and the global maximum is again found at $C^\ast$, but varying 
around a mean value near $1/e$. (The transition from $C$, i.\,e.\ the
expected value, to $C^\ast$, the actual mean value of the random probe,
slightly reduces this value.\footnote{Random probe mean values are known to be 
opportunist. By definition they are minimizing the
variance, and therefore do also decrease the measure of spread 
defined here.} This fact will be investigated later.) 
Hence one has 
\begin{eqnarray}
\lefteqn{\cm(\cf) := \int\limits_0^{C^\ast}
  \left(1-\frac{\Delta t}{C^\ast}\right)
  \cdot \cf(\Delta t)\,d \Delta t}
\label{tdef} \\
\nonumber\lefteqn{
\mbox{(with $C^\ast = \int \Delta t \,\cf(\Delta t)\,d\Delta t$)}}
\end{eqnarray}
being a functional on the set of all pdf defined on 
$[0,\infty[$, with the property $\cm (f_C)= 1/e$.

Provided $C^\ast$ is already computed, $\cm$ can be calculated
from the pdf of the $\Delta T_i$ between $0$ and $C^\ast$ alone. It will be
higher than expected if untypical excesses of small $\Delta T_i$ 
are present (i.\,e.\ in the case of burst-like phenomena), and 
lower than expected for untypical excesses of $\Delta T_i$ near 
$C^\ast$ (i.\,e.\ for untypically regular temporal structures compared
to the average Poisson process).

Fixing the extreme cases  
\begin{enumerate}
\item the heartbeat function\\ $\heartsuit (\Delta t) := 
      \delta(\Delta t - C^\ast)$\\
      belonging to the case of all $\Delta T_i$ being equal, leading to 
      $\cm (\heartsuit )=0$, and
\item the needle peak function\\ $\spadesuit (\Delta t) := 
      \lim_{\varepsilon \to 0}\,
      (1-\varepsilon)\cdot \delta (\Delta t) + \varepsilon 
      \cdot\delta (\Delta t - C^\ast/\varepsilon))$ \\
      characterizing the case of almost all $\Delta T_i=0$,
      and the rest (e.\,g.\ one of them) being huge, resulting 
      in the mean value of $C^\ast$, which leads to $\cm (\spadesuit )\to 1$.
\end{enumerate}
it can be noted for any  pdf on $[0,\infty[$ with a mean value of $C^\ast$:
\be
  \cm (\cf) \in [0,1[ 
\ee

From the above it can be concluded that $\cm$ defines a well-behaved
and natural measure of variability, therefore $\cm$ is chosen
to be the measure of the exp-test. It should be stressed that
only the spectrum of time intervals {\em between} two consecutive events
is entering, thus giving this test a rather differential character.

It remains the derivation of the actual distribution function for the
$\cm(\cf)$ for given $N$ (in the following denoted $\cm_N$ distribution)
under the zero hypothesis. In any case, this pdf is independent on
the particular value of $C^\ast$, since this constant only scales the time,
whereas $\cm$ is a dimensionless quantity. Analytically exact solutions are
\begin{figure}[b!]
  \begin{flushleft}
  \epsfig{file=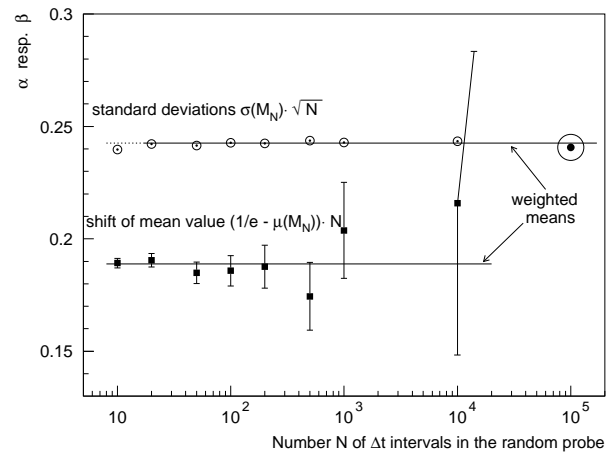,width=8cm}
  \vspace{-0.3cm}
  \caption{\label{t0mom} The standard deviations $\sigma(\cm_N)\cdot\sqrt{N}$ 
           and the shift of the mean value $(1/e - \mu(\cm_N))\cdot N$
           of the $\cm_N$ distribution from \mc\ calculations. Shown are
           also the weighted mean values (straight lines). The errors for
           the spread values (circles) are indicated by the size of the
           central spot. Sample sizes are $1.3\cdot10^5$ for each tested value
           of $N$ (except for $N=10^5$ with a sample size of $1.3\cdot10^4$).}
  \end{flushleft}
\end{figure}
\begin{figure*}[t!]
  \begin{flushleft}
  \epsfig{file=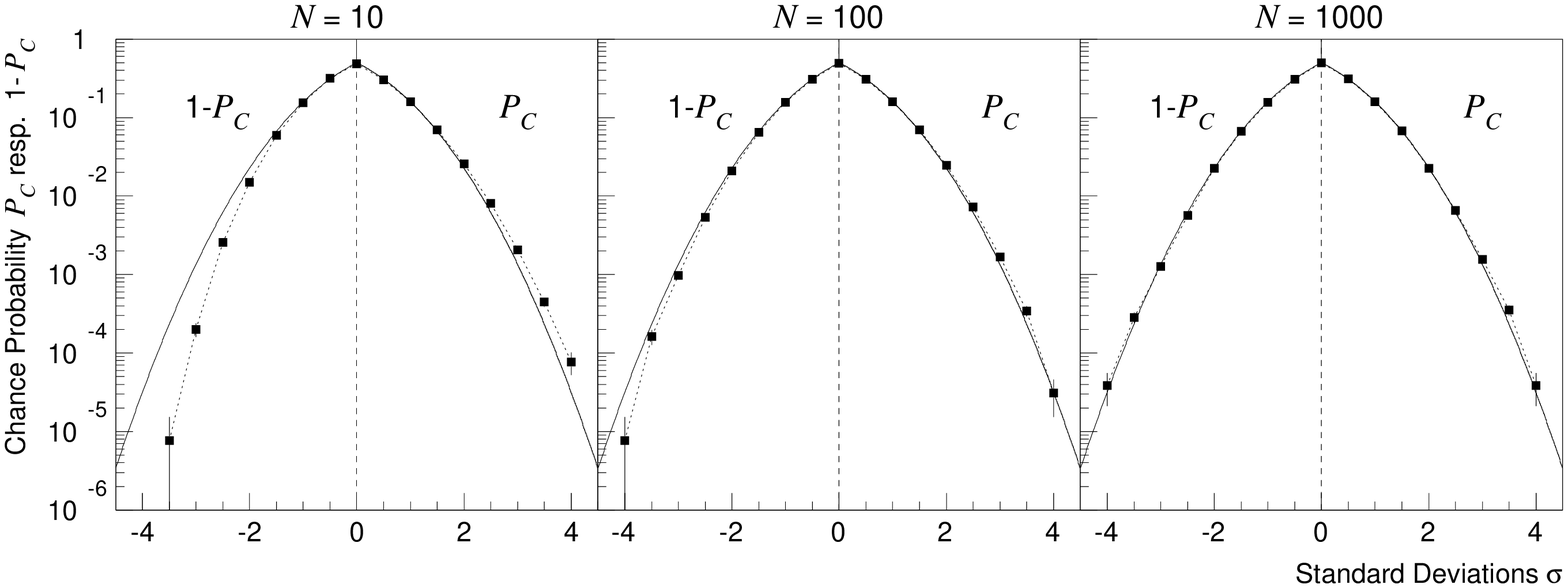,width=17cm}
  \caption{\label{normvpl} Comparison of the $\cm_N$ distribution
           with the normal distribution $N(1/e-\alpha/N,\beta/\sqrt{N})$:
           shown are the chance probabilities from
           the normal distribution (solid lines) and the relative
           frequencies from \mc\ calculations for $\cm_N$ (symbols and
           dashed lines) against $\sigma$. Each sample distribution consists of
           $1.3\cdot10^5$ events.}
  \end{flushleft}
\end{figure*}
surely possible but not attempted here, because a considerable
expenditure has to be expected. Instead, a semi-analytical approach is
followed, using the fact that all spreads occuring here are resulting 
from spreads in the polynomial distribution, all of them scaling with
$\sqrt{N}$, thus leading for dimensionless variables such as $\cm$
to asymptotic standard deviations $\sigma\propto 1/\sqrt{N}$. 
For a similar reason,
the difference of the mean value $\mu$ and $1/e$ will be, again at least
asymtotically, proportional to $1/N$. 
Furthermore it can be expected, according to the central
limit theorem of statistics,
that for increasing $N$ the distribution as a whole will tend towards a normal
distribution.
Therefore only two constants $\alpha,\,\beta$ need to be determined, for
which the following asymptotic scaling laws apply:
\be \frac{1}{e} - \mu(\cm_N) = \frac{\alpha}{N} \qquad \mathrm{and} \qquad
     \sigma(\cm_N) = \frac{\beta}{\sqrt{N}}
\label{abdef}
\ee
In simulations of Poisson processes by generating time sequences with the
help of a pseudo random number generator, there could not be found 
any significant deviations from \gl{abdef} for $N$ ranging between
$20$ and $10^5$ (using $1.3\cdot10^5$ individually generated sequences
per considered number of $N$, see \,\ab{t0mom}). With weighted means of the
obtained values, $\alpha,\,\beta$ could be constrained to 
\be
  \alpha = 0.189 \pm 0.004 \qquad\qquad \beta = 0.2427 \pm 0.0002
\label{albeta}
\ee
There is only one faint (but significant) deviation from the $\sqrt{N}$
scaling law of the spread at $N=10$ (the actually obtained number there is 
$\sigma\cdot\sqrt{10}=0.2400\pm0.0005$). However, the use of the scaling
law at $N=10$ thus corresponds to an error of $\approx1\%$ on a significance
scale, which is negligible in almost all cases. 


After having determined the mean values and the standard deviations 
with sufficiently high precision for most purposes, 
only the shape of the $\cm_N$ distribution has to be compared to that of a 
normal distribution $N(1/e-\alpha/N,\beta/\sqrt{N})$ with the same
mean and variance. In \ab{normvpl}\ the generated relative event frequencies are
compared to $N(1/e-\alpha/N,\beta/\sqrt{N})$-distributed probabilities,
for examples of $N=10,\,100,\,1000$. Due to the fact that 
$\mu(\cm_N)\approx1/e$ is not in the middle of the allowed
interval ($[0,1[$) for $\cm_N$, a faint positive skewness of the
$\cm_N$ distribution can be noticed for smaller $N$. The peculiar drop
of frequencies towards larger negative significances for $N=10$ results from
this, too, since $\cm_{10}=0$ corresponds to $\approx -4.5\,\sigma$.
Such an effect is already hardly noticeable for $N=20$. 
One notes that already for $N=10$ and
moderate positive deviations from the expectation value 
($<\!2\sigma$) the actual $\cm_N$ distribution
follows the normal distribution with a good precision,
whereas for $N=100$ the approximation with the normal distribution is
globally valid for most practical cases.
So only for extremely high precision requirements 
individual \mc\ calculations are necessary to assess the exact significance
of an effect.
\section{Using background events for the exp-test}
The exp-test as developed so far can be 
readily applied in data analyses of experiments
that have a sufficiently constant temporal acceptance. In the case of 
a varying acceptance function \gl{dpoisson}\ is not valid, as one cannot
expect a Poisson process any longer. (This case is sometimes referred to
as {\em inhomogenous Poisson process}, see Chatfield \cite{chatf}.)
 
If the temporal acceptance is discretely switched on and off, the problem
can be easily solved by excluding the time intervals during which the
data aquisition was actually switched off or blocked (e.\,g.\ due to
electronics dead time). 
In the general case (continuously varying acceptance $a(t)$, which is the
typical case in earth-bound $\gamma$-ray experiments), the variable 
$t$ has to be replaced by a scaled effective time $\tau$ 
($a(t)\cdot dt \longrightarrow d\tau$), in order to apply the exp-test
as defined above, but this would require a
precise and complete knowledge of the temporal acceptance function.
However, a simple reasoning shows that background event times
$(t_{\scri{BG},j})$ (being from a Poisson process in $\tau$ as well),
simultaneously collected with the event times $(t_i)$ of interest, 
could serve the same purpose if the background has a sufficiently high
temporal density (e.\,g., in the case of cosmic rays, if the background
events stem from a solid angle which is large compared to the one
that defines the ``on-source" events). 
In this case no explicit knowledge about $a(t)$ is required any longer.
Instead, one can use the distribution of the $\Delta N_{\scri{BG}}$ between
two consecutive on-source events directly, thus testing, so to speak,
whether the two distributions are mutually Poisson distributed.

Proceeding for this purpose from the pdf of the generalized time intervals
$\Delta \tau$ (see \gl{dpoisson}) between two on-source events
(the actual acceptance function $a(t)$ does not play a role in the end;
the only exploited feature is that it scales the on-source and the
background distributions simultaneously):
\be
  f_{\scri{on}}(\Delta \tau) =
       \frac{1}{C_{\scri{on}}} 
           \exp\left(-\frac{\Delta \tau}{C_{\scri{on}}}\right)
\label{aexpdis}
\ee
with a certain, fixed $ C_{\scri{on}}$.\\
Now let the background events be from a Poisson process as well,
with an expectation value $\lambda_{\scri{BG}}$ in a generalized time
interval $\Delta\tau$ of
$$ 
  \lambda_{\scri{BG}}(\Delta\tau) = \Delta \tau / C_{\scri{BG}}
$$
From \gl{aexpdis}\ follows the distribution of the
{\em mean values} for the number of background events between two
on-source events:
\be
f(\lambda_{\scri{BG}}) = \frac{1}{C}\cdot 
       \exp\left(-\frac{\lambda_{\scri{BG}}}{C}\right) 
\label{lambdis}
\ee
(with a global expectation value of $C:= C_{\scri{on}}/C_{\scri{BG}}$)\\
The complete pdf for the {\em number of background events between two
consecutive on-source events} (in the following called ``Inter-Events")
can be derived now as a weighted (with
$f(\lambda_{\scri{BG}})$)  integration of the
Poisson distribution $P_{\lambda_{\scri{BG}}}$ (see \gl{poisson})
over $\lambda_{\scri{BG}}$:
\begin{eqnarray}
\nonumber
w_C(n) &=& \int\limits_{\lambda_{\scri{BG}}=0}^{\infty}
P_{\lambda_{\scri{BG}}}(n)\cdot
           f(\lambda_{\scri{BG}}) d\lambda_{\scri{BG}}\\
     &= &\frac{1}{C+1}\left(\frac{C}{C+1}\right)^{n}
\label{wn}
\end{eqnarray}
i.\,e.\ a discrete exponential distribution. Expressed with the help of the 
exponential function one yields
\be
w_C(n) = \frac{1}{C+1} \cdot \exp\left(
       -\log\left(1+\frac{1}{C}\right) \cdot
n\right)
\label{wne}
\ee
(compare\ \gl{dpoisson}). Thus the number of background events
can be used as a substitute for a generalized clock (measuring 
$\tau$ intervals). The precision of this ``clock" will be defined
by $C$ (i.\ e.\ the ratio of the number of background events to the
number of time intervals to test, or, equivalent, the mean value of all
Inter-Events).

Analogously to the definitions in the last section let there be a
frequency distribution $\dw(n)$ of Inter-Events from a random probe
of $N$ intervals, with a mean value of 
\be
  \sum_{n=0}^\infty n\cdot \dw(n)\, = \,C^\ast
\label{castd}
\ee
Eqs.\,\ref{poismom1},\,\ref{poismom2} are valid analogously, and one 
writes
\begin{eqnarray}
 \lefteqn{ \dh(K):= \sum_{n=0}^K\left(1-\frac{n}{C}\right)w_C(n) =
  \frac{K+1}{C+1}\left(\frac{C}{C+1}\right)^{K}}\\
\nonumber\lefteqn{
  \mathrm{with}\quad
  \dh(-1) = \dh(\infty) = 0}
\end{eqnarray}
with the maximum found at $K=[C]$.\footnote{Here and in the
following the pair of square brackets $[\ldots]$ denotes the
integer function.}
The value at $[C]$ depends obviously on $C$ itself, but tends again to
$1/e$ if $C\to\infty$. With 
\be
  \dm(\dw) := \sum_{n=0}^{[C^\ast]}
  \left(1-\frac{n}{C^\ast}\right)
  \cdot \dw(n)\quad\enspace(\,C^\ast \mbox{\,from \gl{castd}})
\label{tdefd}
\ee 
again $\dm(\dw)\in [0,1[$, and for $\dw(n)$ from Inter-Events
in the case of a Poisson process, that $\dm(\dw)$ has a mean value near
\be
  \dm_0(C^\ast) = \frac{[C^\ast]+1}{C^\ast+1}\cdot
  \left(\frac{C^\ast}{C^\ast+1}\right)^{[C^\ast]}
\label{meandis}
\ee 
and a variance that depends this time also on $C^\ast$.

Naming the mean value resp.\ the standard deviation 
of the corresponding $\dm_{C^\ast,N}$ distribution
$\mu_{C^\ast}(N)$ and $\sigma_{C^\ast}(N)$, respectively, and defining 
(analogously to \gl{abdef})
\be \dm_0(C^\ast) - \mu_{C^\ast}(N) = \frac{\alpha_{C^\ast}}{N}
  \qquad\qquad
     \sigma_{C^\ast}(N) = \frac{\beta_{C^\ast}}{\sqrt{N}}
\label{abdefd}
\ee
then empirically the following parametrization can be found:
\begin{eqnarray}
  &\alpha_{C^\ast}\approx \alpha\cdot k_1^{\frac{1}{C^\ast+k_2}}
  &\qquad\quad
  \beta_{C^\ast}\approx \beta\cdot k_1^{\frac{1}{C^\ast+k_2}} \\
\nonumber
  \mathrm{with}\quad & k_1 = 1.67 \pm 0.02
  & \qquad\quad 
   k_2 = 0.37 \pm 0.03
\label{k1k2}
\end{eqnarray}
($\alpha$, $\beta$ from \gl{albeta})\\
The least squares fit that leads to $k_1$ and $k_2$ is shown in
\begin{figure}[!t]
  \begin{flushleft}
  \epsfig{file=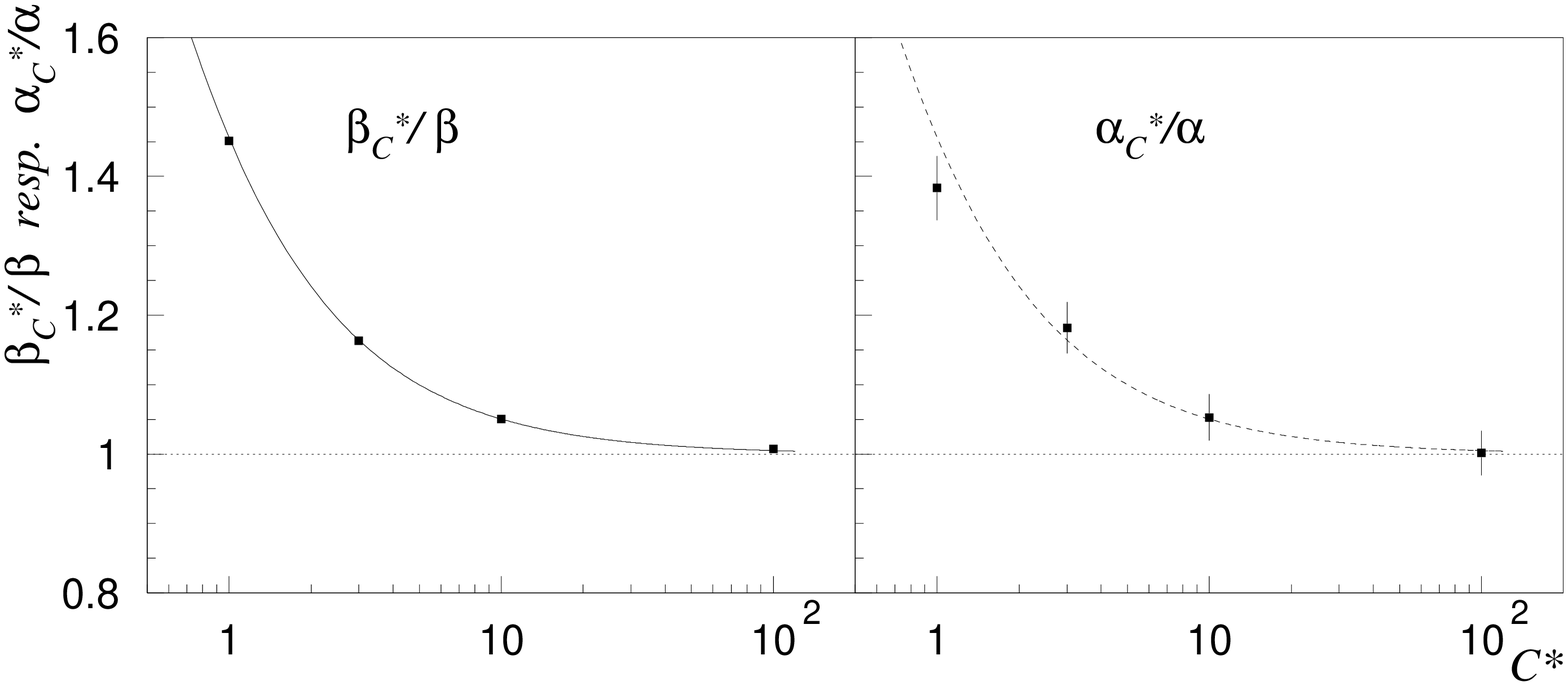,width=8.8cm}
  \caption{\label{t0cdis} The dependence of $\beta_{C^\ast}$
   and $\alpha_{C^\ast}$ from $C^\ast$ from \mc\ calculations (see text).
    The solid line
   (left) shows the least squares fit (repeated with dashes at the right
   hand side). The sample sizes for each considered $C^\ast$ are $1.8\cdot10^5$.}
  \end{flushleft}
\end{figure}
 \ab{t0cdis}\ (left). 
Since the spreads can by far more accurately be assessed than the mean values,
again the fit to them provides the more precise numbers. 
However, in \ab{t0cdis}\ (right) it can be seen that the dependency of
$\alpha_{C^\ast}/\alpha$ on $C^\ast$ is compatible with the same
parametrization, so that the same functional behaviour may well be assumed.
Summarizing, it can be stated for the discrete case that was considered here,
that the $\dm_{C^\ast,N}$ distribution equals approximately a normal
distribution $N(\dm_0(C^\ast)-\alpha_{C^\ast}/N,\beta_{C^\ast}/\sqrt{N})$,
with only small deviations similar to the ones shown in \ab{normvpl}. For
$C^\ast \to \infty$ the distribution of $\dm_{C^\ast,N}$ tends in all
aspects towards the $\cm_N$ distribution. Since decreasing $C^\ast$
lead to increasing spreads $\beta_{C^\ast}$, for optimization of the 
sensitivity $C^\ast$ should be chosen as large as possible.

As a last general remark that applies to the continuous case as well as
to the discrete case it should be stressed that the measure of the exp-test
$\cm(\cf)$ or $\dm(\dw)$ can be assessed with running
sums when one allows for two sequential passes over the data: in the first
pass, $C^\ast$ can be determined, whereas in the second pass the sum in 
\gl{tdef}\ or \gl{tdefd}\ can be calculated. This possibility is a major
advantage in comparison to binned procedures.
\section{\label{ptest3} Sensitivity of the exp-test}
This section deals with an application of the exp-test in a general
scenario, in order to characterize its sensitivity.

Consider the case of a sporadically active source 
with a duty cycle $q\in ]0,1]$, during which it ``pollutes" the undisturbed
background with additional events from another Poisson process (i.\,e., the
constellation of the first question mentioned in the introduction).
Let the total number of registered events be $N$, the number
of background events be $N_1$, and the number of additional events
from the source be $N_2$ (i.\,e.\  $N = N_1 + N_2$). 
Without loss of generality, the total observation time is set to $1$
(see sketch in \ab{dutyplot}).
The same pattern applies to the case of undiluted source events in a search
for a signal of variablility (second question posed in the introduction),
in which case $N_1$ denotes the quiescent flux 
$\Phi_{\mathrm{low}}$ and $N_2/q+N_1$ corresponds to the flux
$\Phi_{\mathrm{high}}$ during the active state of an object.

To begin with the consideration of the continuous $\cm_N$ statistics,
it follows for the pdf $\tilde{F}$ of $\Delta t$ to be
\begin{eqnarray}
\lefteqn{\tilde{F}(\Delta t) = \frac{N_1}{N}(1-q)\cdot f_{C_1}(\Delta t) +
                \frac{N_2+q N_1}{N}\cdot f_{C_2}(\Delta t)} \\
\nonumber\lefteqn{  \mathrm{with} \quad C_1=\frac{N}{N_1} \enspace
         \mbox{,}\enspace C_2=\frac{qN}{N_2 + q N_1}} \\
\nonumber\lefteqn{ \mathrm{and}\quad f_{C_i}(\Delta t)
\,\,\mbox{from\,\,\gl{dpoisson}}}
\label{fduty}
\end{eqnarray}
(i.\,e., according to \gl{poismom2}, $C^\ast = 1$, without loss of
generality)\\ 
From \gl{tdef}\ it follows an expectation value for $\cm(\tilde{F})$ of
\be
   \langle\cm(\tilde{F})\rangle = \frac{1}{e}\exp\left(\frac{N_2}{N}\right)
        \cdot\left(
        1 -
        q + q\exp\left(-\frac{N_2}{q N}\right)\right)
\label{t0cexp}
\ee
It is important to note that, due to the differential nature of the test,
$\tilde{F}(\Delta t)$ and consequently $\langle\cm(\tilde{F})\rangle$
apply not only to a flux pattern as shown in \ab{dutyplot}, but to 
{\em all} two-leveled light curves. More specifically, 
they are neither affected by
the actual position of the time window of the active state, nor by an
arbitrary number of interrupts of the active state, as long as the total
duty cycle is kept fixed and $N_2$ is large compared to the number of
interrupts.\\[4pt]

\begin{figure}[!t]
  \begin{flushleft}
  \epsfig{file=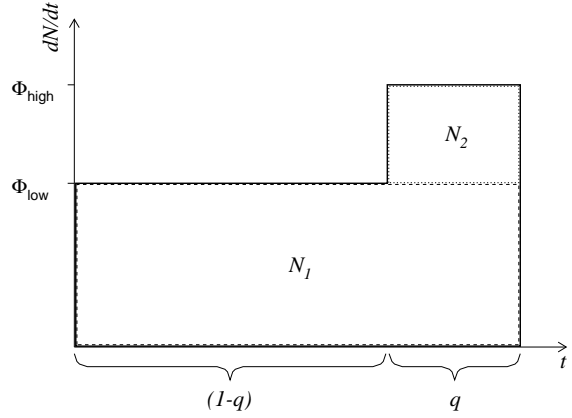,width=7.5cm}
  \caption{\label{dutyplot} Sketch of the mean event rates from a source 
  with a variable flux (two flux levels).}
  \end{flushleft}
\end{figure}
In the notion of the 1.\ question, one will ask for the {\em significance}
of the exp-test result in comparison to the significance of a DC excess.
For the exp-test, a significance 
$S\sc{exp}$ of \footnote{In this consideration
of expectation values there is no influence from the shift of the mean value
$\alpha/N$. In any case of real random probes with finite $N$, this shift
cancels out since it concerns the actual probe as well as the expectation.}
\be
\langle S\sc{exp}\rangle\simeq
    \sqrt{N}(\langle\cm(\tilde{F})\rangle - 1/e)/\beta
\label{cexpsig}
\ee
can be expected. For any fixed $N_2/N$, $\langle S\sc{exp}\rangle$ is a positive, strictly
monotone convex falling function of $q$, being zero at $q=1$. 
In the limit of an arbitrary but fixed ratio $N_2/N \ll 1$
one gets from Eqs.\ \ref{t0cexp},\,\ref{cexpsig} 
\be
\langle S\sc{exp}\rangle =\frac{1}{e\beta}\cdot
\left\{ 1-\frac{N}{N_2}q\left( 1-\exp\left(-\frac{N_2}{qN}\right)\right)
\right\}\cdot\frac{N_2}{\sqrt{N}}
\ee
Under the above limit the expectation for the DC significance
$S\sc{DC}$ in the case of a perfect knowledge of the background level is
\be
\langle S_{\mathrm{DC}}\rangle\simeq N_2/\sqrt{N}
\label{cdcsig}
\ee
Thus
\begin{eqnarray}
\nonumber\lefteqn{\langle S\sc{exp}\rangle =}\\
& &\enspace\underbrace{\frac{1}{e\beta}}_{\textstyle\approx 1.51}\cdot
\left\{ 1-\frac{N}{N_2}q\left( 1-\exp\left(-\frac{N_2}{qN}\right)\right)
\right\}\cdot\langle S\sc{DC}\rangle \label{csig2}\label{critnaeh}
\end{eqnarray}
i.\,e.\ in the limit $q\to 0$ (that means extremely burst-like behaviour
of the source) one has to expect a significance from the exp-test
that is 50\% higher than the DC result, so it can be stated that it is worth
to apply the exp-test in the case when some DC excess is present from an
appropriate candidate source. It should be emphasized once again
that both significances are perfectly independent if the zero hypothesis 
is true.

Obeying the limit from above and defining $q\sc{crit}:=N_2/N$, for 
$q\ll q\sc{crit}$ it follows from \gl{csig2} that
\be 
 \langle S\sc{exp}\rangle = \frac{1}{e\beta}\cdot \left( 1-
  \frac{q}{q\sc{crit}}\right)\cdot \langle S\sc{DC}\rangle
\label{clin}
\ee 
while for $q=q\sc{crit}$ one obtains
\be
 \langle S\sc{exp}\rangle\Big|_{q=q\sc{crit}} = \frac{1}{e}
\langle S\sc{exp}\rangle\Big|_{q=0}
\ee
i.\,e.\ in the case of moderate DC significances, only for 
$q\lesssim q\sc{crit}$ a significance from the exp-test that is worth
mentioning can be expected. In the context of the first question, the exp-test
therefore should be regarded as a test on burst-like temporal structures.

In the case of the discrete $\dm_{C^\ast,N}$ statistics,
qualitatively the same signature for the considered scenario occurs.
One finds for the probability distribution 
\begin{eqnarray}
\lefteqn{
  \widetilde{\dw}(n) = \frac{N_1}{N}(1-q)\cdot w_{C_1}(n) +
                \frac{N_2+q N_1}{N}\cdot w_{C_2}(n)}\\
 \nonumber\lefteqn{\mathrm{with} \quad  C_1=C^\ast\cdot\frac{N}{N_1} \enspace
         \mathrm{,} \enspace C_2=C^\ast\cdot\frac{qN}{N_2 + q N_1}}\\
 \nonumber
 \lefteqn{ \mathrm{and}\quad w_{C_i}(n)\,\, \mbox{from\,\,\gl{wn}}}
\label{fdutyd}
\end{eqnarray}
For arbitrary $C^\ast$ it is not possible to give a closed expression
for $\langle\dm(\widetilde{\dw})\rangle$ analogously to \gl{t0cexp}.  
However, the main feature of the discrete case can be assessed as
follows.
Using this time
\be
\langle S\sc{exp}\rangle\simeq
    \sqrt{N}(\langle\dm(\widetilde{\dw})\rangle - 
    \dm_0(C^\ast))/\beta_{C^\ast}
\label{dexpsig}
\ee
one finds after lengthy 
but trivial calculations for integral $C^\ast$ and $q \ll N_2/N\ll 1$:
\be
\langle S\sc{exp}\rangle \approx 
 \left(\frac{C^\ast}{C^\ast+1}\right)^{C^\ast+1}\cdot
 \frac{1}{\beta_{C^\ast}}\cdot\left(
  1-\frac{q}{q\sc{crit}}\right)
  \cdot\langle S\sc{DC}\rangle
\label{t0dexp}
\ee
Under the limitations stated above this yields for $C^\ast=1$ 
\be
\langle S_{\mathrm{exp}}\rangle \approx 
   \underbrace{\frac{1}{4\beta_1}}_{\textstyle\approx0.7}\cdot 
   \left(1-\frac{q}{q\sc{crit}}\right)\cdot 
   \langle S_{\mathrm{DC}}\rangle
\ee
but for $C^\ast=10$ already
\be
\langle S\sc{exp}\rangle \approx 
\underbrace{\left(\frac{10}{11}\right)^{11}\cdot\frac{1}{\beta_{10}}}_{
   \textstyle\approx1.4}
  \cdot \left(1-\frac{q}{q\sc{crit}}\right)\cdot 
    \langle S_{\mathrm{DC}}\rangle
\ee
Please note in this context, that for the calculation of the DC significances
a perfect knowledge of the background level was still assumed.

The overall shape of the function $\langle S\sc{exp}\rangle/\langle
S_{\mathrm{DC}}\rangle$ in dependence of $q$ is, apart from the lower
absolute values, very similar to \gl{critnaeh}. It therefore can be
summarized that the use of background events qualitatively yields the same
results as the continuous $\cm_N$ statistics, while the quantitative loss
in sensitivity in dependence of $C^\ast$ can be obtained from a comparison
of \gl{clin}\ and \gl{t0dexp}.\\[4pt]

Turning now to the second question mentioned in the introduction
(pure source events), one will be interested in the sensitivity
to detect a certain flux ratio $\Phi\sc{high}/\Phi\sc{low}$
when a duty cycle $q$ is given, or vice versa. This shall only be
carried out here for the continuous $\cm_N$ statistics.

Defining $r:=\Phi\sc{high}/\Phi\sc{low}$, one gets
from \gl{t0cexp}:
\begin{eqnarray}
   \nonumber\lefteqn{
   \langle S\sc{exp} \rangle =  
   \frac{\sqrt{N}}{\beta}\cdot 
   \left(\langle\cm(\tilde{F})\rangle - \frac{1}{e} \right)} \\
  \nonumber & &= 
   \frac{\sqrt{N}}{e\beta}\cdot \bigg\{
   \left( 1-q+q\cdot\exp\left(
      -\frac{1-r}
            {q+(1-q)r} \right)\right)\\
   & & \qquad\qquad\qquad\qquad\qquad\cdot \exp\left(
       \frac{q -q\cdot r}
            {q+(1-q)r} \right)
      - 1 \bigg\}
\label{dutysig}
\end{eqnarray}
In \ab{varflux} the resulting expectations for the significances are
displayed as contour lines in the $q$-$r$ plane.
\begin{figure}[!t]
  \begin{flushleft}
  \epsfig{file=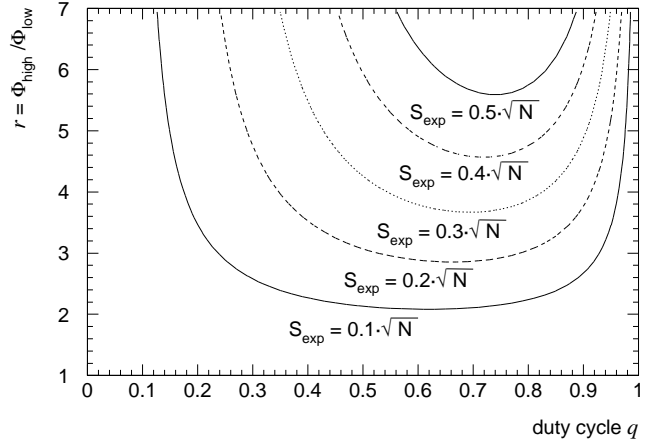,width=8.5cm}
  \caption{\label{varflux} Contour lines of the exp-test significance
  $S\sc{exp}$ in the $q$-$r$ plane. This figure allows for the
  determination of the detectable $q$-$r$ region if the total number
  of events $N$ and a required minimum value of $S\sc{exp}$ are given.}
  \end{flushleft}
\end{figure}
Making use of the fact that the the assumed two-level model is most
efficient to produce an effect on $\cm(\tilde{F})$ with respect to an
upper limit of the ratio of maximum to minimum flux levels present for
a given source, for a sufficiently high significance $S\sc{exp}$ it is
immediately possible to deduce a {\em lower limit} of actual flux variations
$\Phi\sc{max}/\Phi\sc{min}$ for a source. For the given example values
of $S\sc{exp}$, these ratios can be read off
from \ab{varflux}\ as the ordinate values of the cusps of the displayed curves.
For any given value of $S\sc{exp}$, $\Phi\sc{max}/\Phi\sc{min}$ has to
be obtained with graphical or numerical methods, regarding \gl{dutysig}\ as
an implicit function $r(q)$ with $S\sc{exp}$ as a parameter.

It can be concluded that in this case, where the significance is expressed
in terms of flux levels in the high and the low states rather than in
terms of event numbers, the exp-test is a quite broad-banded tool
in the search for variabilities with respect to the duty cycle $q$.
\section{Exp-test versus Kolmogorov test: a comparison}
In order to recognize the strengthes of the exp-test on the one hand and
the Kolmogorov test on the other, the latter shall be briefly reviewed here.

The Kolmogorov test uses the maximum $D$ of the absolute differences of the
cdf $G(t)$ of the temporal distribution of the event sequence and the expected
cdf $g(t)$ (i.\,e.\ an equal distribution in the case of uniform temporal
acceptance):
\be
 D:= \max \left| G(t)-g(t)\right|
\ee
Being $N$ the total number of events, it can be shown that the chance 
probabilities $P\sc{kolmo}$ in dependence from $D$ under the zero hypothesis 
are asymptotically distributed according to Kolmogorovs $\lambda$ distribution:
\be
 P\left(D>\frac{\lambda}{\sqrt{N}}\right) \simeq 2\cdot
 \sum_{k=1}^\infty(-1)^{k+1}\,e^{-2k^2\,\lambda^2}
\label{gkolmo}
\ee
(Kolmogorov \cite{kolmo}).\\
Consider now a distribution on $[0,1]$ similar to the one shown in
\ab{dutyplot}, with one uninterrupted activity phase of the length $q$,
containing $N_2$ excess events, but this time with a specified starting time
$t_0\in[0, 1-q]$. The corresponding cdf is sketched in \ab{kolmog},
and the expected difference $\langle \tilde{D}\rangle$ to the equal
distribution can easily be calculated:
\be
 \langle \tilde{D}_{q,t_0}\rangle = \left(1-q-\min(t_0,1-q-t_0)\right)
    \cdot\frac{N_2}{N}
\ee
Averaging over $t_0$ yields
\be
  \langle \tilde{D}_q\rangle = \frac{3}{4}\left(1-q\right)
    \cdot\frac{N_2}{N}
\ee
Again the question of the {\em significance}\footnote{Calculating
significances for results of the Kolmogorov test is rather unusual, but
shall be performed here for the sake of easier comparison. With
$\mbox{freq}(x) := \left(\sqrt{2\pi}\right)^{-1}\int_{-\infty}^x 
\exp(-z^2/2)\,dz$ every
chance probability $P_C$ corresponds to a significance of 
$S=-\mbox{freq}^{-1}(P_C)$.} one may expect from this constellation shall be
posed. An easy estimate can be obtained under the assumption that $\tilde D$
already determines the actual $D$ to a sufficient precision in the test,
\begin{figure}[t!]
  \begin{flushleft}
  \epsfig{file=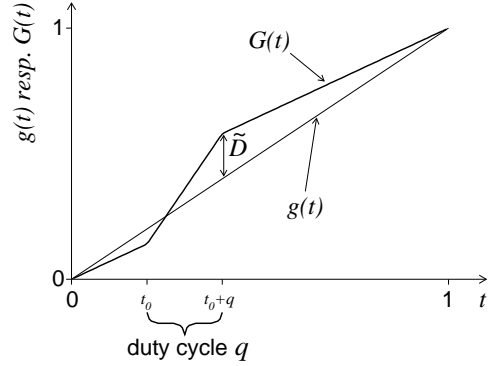,height=5cm}
  \caption{\label{kolmog} Sketch of the cdf $G(t)$ of event times from a
  source being in an active state during $[t_0,t_0+q]$, and the cdf $g(t)$
  of the zero hypothesis (see text).}
  \end{flushleft}
\end{figure}
that means, a difference $D\ge\tilde D$ is fairly improbable under the zero
hypothesis. Under this condition, in the series in
\gl{gkolmo}\ all but the first term are negligible, and it follows
\be
 P\sc{kolmo}\left(D>\tilde{D}_q\right) \approx 2\cdot e^{-\frac{9}{8}
   (1-q)^2\frac{N_2^2}{N}}
\label{pkolmolim}
\ee
From
\begin{eqnarray}
\nonumber\lefteqn{\frac{1}{\sqrt{2\pi}}\cdot\frac{1}{x}\cdot
   e^{-\frac{1}{2}x^2} =
   \frac{1}{\sqrt{2\pi}}\cdot\int\limits_x^\infty
      \left(1+\frac{1}{z^2}\right)\cdot e^{-\frac{1}{2}z^2}\,dz}\\
  & & \qquad = \left(1+\mathcal{O}\left(\frac{1}{x^2}\right)\right)\cdot
      \underbrace{\frac{1}{\sqrt{2\pi}}\cdot\int\limits_x^\infty   
       e^{-\frac{1}{2}z^2}\,dz}_{P_C(x)}
\end{eqnarray}
a first order asymptotic expansion of the standard normal
distribution can be deduced:
\be
  P(S)\simeq \frac{1}{\sqrt{2\pi}}\cdot \frac{1}{S} \cdot
  e^{-\frac{1}{2}S^2}
\ee
from which one obtains in a comparison with \gl{pkolmolim} 
for the significance $S\sc{kolmo}$ of the Kolmogorov test result
\be
  \langle S\sc{kolmo}\rangle \approx \frac{3}{2} (1-q) \frac{N_2}{\sqrt{N}}
  \approx \frac{3}{2} (1-q) \langle S\sc{dc}\rangle
\label{kolapprox}
\ee
Numerical calculations show that this estimation is indeed approached for
large $N$ and $\langle S\sc{dc}\rangle$, while for small $N$ and 
$\langle S\sc{dc}\rangle$ $\langle S\sc{kolmo}\rangle$ is somewhat smaller
(Prahl \cite{prahl}).
For an example of $N=100$ and $\langle S\sc{DC}\rangle=3\,\sigma$  
the significance $\langle S\sc{kolmo}\rangle$ is reduced by $\approx20\%$ for
$q\to0$. Nonetheless, \gl{kolapprox}\ may serve as a rough estimate
for all practical purposes. 

It is obvious that for splitted activity intervals $\langle S\sc{kolmo}\rangle$
will be much lower. In a comparison with the exp-test (see \gl{critnaeh})
therefore it has to be stated 
that for {\em one single outburst} of a source with $q\ll N_2/N$
the expected significances from the Kolmogorov test and
from the exp-test are similar (but slightly higher from the latter),
while for a comparably long and uninterrupted 
activity interval the Kolmogorov test is more sensitive, whereas the exp-test 
is better suited to find some or many short outbursts.

Finally, the correlation of both tests under the zero hypothesis has to
be studied. 
\begin{figure}[tb]
  \begin{flushleft}
  \epsfig{file=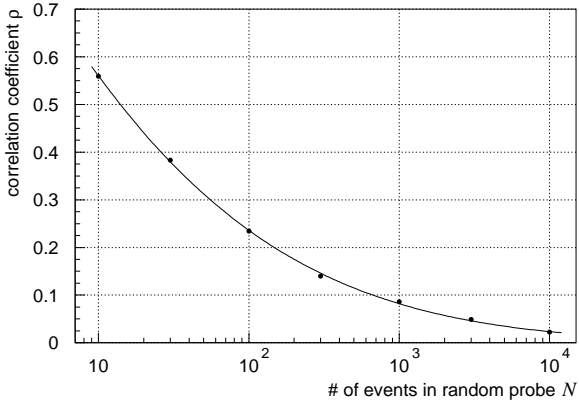,height=5.5cm}
  \caption{\label{corrc} The correlation coefficient between the
  significance result
  of the exp-test and that of the Kolmogorov test for \mc\ generated Poisson
  processes, versus the number $N$ of tested events. The sample size for each
  considered value of $N$ is $1\cdot10^5$.}
  \end{flushleft}
\end{figure}
This is again performed in \mc\ calculations. It turns out that the 
correlation coefficient $\varrho$ between $S\sc{kolmo}$ and 
$S\sc{exp}$ is a monotonically decreasing function of
$N$, being $\varrho\approx 0.56$ for $N=10$, but already only 
$\varrho\approx 0.23$
for $N=100$. The graph obtained for $\varrho(N)$ is shown in \ab{corrc}.
One concludes that it is worth to apply both tests in a search of 
variable sources, maybe except for {\em very small} event numbers.

It should be mentioned that, like for the exp-test, the result of the
Kolmogorov test is completely independent from the DC test result if the
zero hypothesis is valid. Furthermore, the application of Smirnovs variant
(Smirnov \cite{smirn}) allows for the use of simultaneously acquired
background events to define the zero hypothesis in the case of nonuniform
temporal acceptance. Just like for the exp-test, the Kolmogorov test result
can also be determined from runnig sums during two sequential passes over the
experimental data.
\section{Summary}
%
It has been shown that, given $N$ time intervals 
$\lbrace\Delta T_k\rbrace _{k=1\ldots N}$ between each pair of consecutive
event times of the sequence to test (resp.\ $N$ Inter-Events 
$\lbrace n_k\rbrace_{k=1\ldots N}$ in the discrete case), the quantity 
\begin{eqnarray}\displaystyle
 \nonumber & \cm = \frac{1}{N}\sum\limits_{T_k<C^\ast} \left(1-\frac{T_k}{C^\ast}\right)\\
  \displaystyle\nonumber &\left(\enspace \mathrm{resp.}\quad \dm = \frac{1}{N}
  \sum\limits_{n_k<C^\ast}
  \left(1-\frac{n_k}{C^\ast}\right) \quad \right)
\end{eqnarray}
is asymptotically normal distributed according to
$N(1/e-\alpha/N,\beta/\sqrt{N})$
(resp.\ according to $N(\dm_0(C^\ast)-\alpha_{C^\ast}/N,\beta_{C^\ast}/\sqrt{N})$)
if the temporal sequence represents a Poisson process
(resp. inhomogenous Poisson process).
The distribution of the time intervals $\cf(\Delta t)$
resp.\ of the Inter-Events $\dw(n)$ is easily accessible, and $\cm(\cf)$
resp.\ $\dm(\dw)$ are quickly calculable from it. In contrast to binned tests,
the measure of the exp-test can be obtained from running sums when
processing the data. The determination of the significance of the exp-test
$S\sc{exp}$ therefore is almost as easy as the calculation of
a DC significance.

In observations that are limited by background fluctuations,
obtained positive significances $S\sc{exp}$ for a tested temporal sequence 
correspond to small
chance probabilities and can directly be interpreted as significances for
burst-like behaviour. For sufficiently small duty cycles, the
significances from the exp-test are expected to be $1.5$ times higher than
the DC significances. The direct use of background events acquired
simultaneously matches perfectly the requirements of typical
experiments in the earthbound $\gamma$-ray astronomy.

Asking for signals of variability in pure source events,
the result of the exp-test represents a simple and broad-banded measure
for deviations from a constant flux. A positive result of the exp-test
in such situations supplies an immediate information about the minimum
flux variations $\Phi\sc{max}/\Phi\sc{min}$ present in the sample.

The results from the exp-test are by construction independent from a DC excess
or deficit, and unaffected by interrupts of the activity cycle as far as
possible. The test can be applied as soon as the number $N$ of events to test
surmounts $\approx10$. 

Although every test on variability contains arbitrary elements, 
special care was taken to find a nonartificial measure
that is suggested by itself when studying the ele\-men\-ta\-ry statistics of
Poisson processes. From the viewpoint of the author, the exp-test
fulfills this criterion. 

\begin{acknowledgements}
      I wish to thank my colleagues from the HEGRA Collaboration
      for valuable comments on the draft of this paper. 
\end{acknowledgements}

\end{document}